\documentclass[pra,aps,showpacs,twocolumn,floatfix, superscriptaddress]{revtex4-1}
\usepackage{graphicx}
\usepackage[ansinew]{inputenc}
\usepackage{array}
\usepackage{color}
\usepackage{amsmath}
\usepackage{amsxtra}
\usepackage{amstext}
\usepackage{amssymb}
\usepackage{latexsym}
\usepackage{dsfont}

\begin{document}

\title{Antibunching in an optomechanical oscillator}

\author{H.~Seok}
\email{hseok@kongju.ac.kr}
\affiliation{Department of Physics Education, Kongju National University, Gongju 32588, South Korea}
\author{E.~M.~Wright}
\affiliation{College of Optical Sciences and Department of Physics, The University of Arizona, Tucson, Arizona 85721, USA}

\begin{abstract}
We theoretically analyze antibunching of the phonon field in an optomechanical oscillator employing the membrane-in-the-middle geometry.  More specifically, a single-mode mechanical oscillator is quadratically coupled to a single-mode cavity field in the regime in which the cavity dissipation is a dominant source of damping, and adiabatic elimination of the cavity field leads to an effective cubic nonlinearity for the mechanics.  We show analytically in the weak coupling regime that the mechanics displays a chaotic phonon field for small optomechanical cooperativity, whereas an antibunched single-phonon field appears for large optomechanical cooperativity. This opens the door to control of the second-order correlation function of a mechanical oscillator in the weak coupling regime.
\end{abstract}

\maketitle

\section{Introduction}

 Cavity optomechanics is a forefront research field in which the motional degrees of freedom of a mechanical oscillator are coupled to optical fields inside an optical cavity, stemming from the interplay through cavity resonance and  radiation pressure forces~\cite{Review1, Review2, Review3, Milburn_book2}. Recent progress in nano- and micro- fabrication techniques have led to impressive milestones including the cooling of a mechanical oscillator to the motional ground state~\cite{Cooling1, Cooling2}, optomechanically induced transparency~\cite{OMIT1}, coherent coupling of optical and mechanical modes~\cite{Coherent1, Coherent2}, entanglement between optical and mechanical resonators~\cite{Entanglement1}, and optically induced interaction between mechanical oscillators~\cite{Muti_mechancis1}.  Cavity optomechanics has numerous applications such as precision measurement of the position of a mirror allowing for a gravitational wave detection~\cite{Sensitivity,  LIGO1}, a realization of macroscopic quantum objects~\cite{Macroscopic1}, and as a fundamental platform for exploring coupling to other quantum systems~\cite{Hybrid1, Hybrid2, Hybrid3}.  
 
To date almost all experiments and treatments of cavity optomechanics are based on linearized optomechanical interactions in the sense that the interaction is linear in both the field and mechanical variables, and are therefore based on single photon-phonon interactions~\cite{Review1, Review2, Review3, Milburn_book2}.  The intrinsic optomechanical interaction is, however,  nonlinear, which comes to the fore in the single-photon strong coupling regime. The nonlinear nature of the optomechanical interaction gives rise to a variety of features previously explored in nonlinear quantum optics~\cite{Nonlinear_optics}, including photon blockade effects~\cite{Antibunched_photon1}, the generation of non-Gaussian states~\cite{nonGaussian}, and nonclassical antibunched mechanical resonators~\cite{Antibunched_phonon1, Antibunched_phonon2, Antibunched_phonon3, Antibunched_phonon4}.  Since the single-photon radiation pressure is too small to realize the nonlinear strong coupling regime in nanofabricated optomechanical systems, several proposals have studied the possibility of an enhanced optomechanical nonlinearity~\cite{Enhancement1, Enhancement2}, and thus sub-Poissonian phonon field, based on an optomechanical system employing two optical modes in the weak coupling regime~\cite{Antibunched_phonon5}.

In this paper, we theoretically analyze an approach for producing an antibunched phonon field based on the membrane-in-the-middle geometry, and in the weak coupling regime~\cite{Middle1, Middle2}.  In particular, a single-mode mechanical oscillator is quadratically coupled to a single-mode cavity field in the regime where the cavity damping is a dominant source of dissipation, resulting in an effective cubic nonlinearity after adiabatic elimination of the cavity field. We show that the mechanical oscillator is coupled to an effective optical reservoir at zero temperature in addition to its own mechanical heat bath at finite temperature.  To avoid the difficulties that arise from the multiplicative noise that appears from the use of the Heisenberg-Langevin equations, we here employ the Schr\"odinger picture.  Then we demonstrate analytically that the mechanics displays a chaotic phonon field with small multiphoton optomechanical cooperativity, whereas an antibunched single-phonon appears for large multiphoton cooperativity.   

This remainder of this paper is organized as follows: Sec.~\ref{sec:Model System} describes the model system, and Sec.~\ref{sec:Density operator formalism} derives the relevant master equation for the mechanical system. In Sec.~\ref{sec:Results}, we employ the complex $P$ representation to investigate the steady-state behaviors of the mechanical oscillator for both the high and low temperature regimes, and the appearance of antibunching.  Finally Sec.\ref{Conclusions} gives our summary and conclusions.

\section{Model System} \label{sec:Model System}
We consider a membrane-in-the-middle optomechanical system in which the single-mode of an optical resonator is quadratically-coupled to a single mechanical mode of effective mass $m$ and frequency $\omega_m$.  The net Hamiltonian governing the optomechanical system is
\begin{equation}
\hat H = \hat H_{\rm opt} + \hat H_{\rm mech} + \hat H_{\rm om}+ \hat H_{\rm loss}, 
\end{equation}
where
\begin{equation}
\hat H_{\rm opt} = \hbar\omega_c \hat a^\dag \hat a + i\hbar(\eta e^{-i\omega_{L}t}\hat a^\dag- \eta^* e^{i\omega_L t}\hat a) ,
\end{equation}
is the Hamiltonian for the single-mode optical field driven by a monochromatic field of frequency $\omega_{L}$ at pumping rate $\eta$, and 
\begin{equation}
 \hat H_{\rm mech} = \hbar\omega_m \hat b^\dag\hat b ,
\end{equation}
is the Hamiltonian for the free mechanical mode. The optomechanical interaction is given by
\begin{equation}
 \hat H_{\rm om} = \hbar g_0\hat a^\dag \hat a (\hat b+\hat b^\dag)^2,
\end{equation}
where $g_0>0$ is the quadratic single-photon optomechanical coupling coefficient, that we choose as positive to avoid any issues of mechanical instability \cite{Previous_paper}.  Finally, $\hat H_{\rm loss}$ describes the interaction of the cavity field and mechanical modes with their associated reservoirs and accounts for dissipation. 

\section{Density operator formalism}\label{sec:Density operator formalism}
The Heisenberg-Langevin equations of motion for our problem can involve multiplicative quantum noise in the presence of nonlinear interactions~\cite{Previous_paper}.  To circumvent these problems we here explore the dynamics of the optomechanical system in the Schr{\"o}dinger picture since the equation of motion describing the optomechanical system is then strictly linear in the density operator.  The dynamics of the optomechanical system under the influence of thermal fluctuations in the quantum regime can then be described by the master equation~\cite{Meystre_book}
\begin{eqnarray}
 \dot{\tilde \rho} &=&-\frac{i}{\hbar}[\hat H_{\rm opt}+\hat H_{\rm mech}+\hat H_{\rm om}, \tilde \rho]+\frac{\kappa}{2}{\cal D}[\hat a]\tilde\rho \nonumber \\
 &&+\frac{\gamma}{2}{\bar n}_{\rm th}{\cal D}[\hat b^\dag]\tilde\rho+\frac{\gamma}{2}({\bar n}_{\rm th}+1){\cal D}[\hat b]\tilde\rho, 
\end{eqnarray}
where $\tilde \rho$ is the density operator for the combined optomechanical system, and the dissipation terms ${\cal D}[\hat o]\tilde\rho$ are of the standard Lindblad form
\begin{equation}
 {\cal D}[\hat o]\tilde\rho = (2\hat o \tilde\rho \hat o^{\dag}-\hat o^\dag \hat o \tilde \rho- \tilde\rho\hat o^\dag \hat o).
\end{equation}
These account for damping of the cavity field with decay rate $\kappa$ due to the coupling to a zero-temperature optical reservoir, and damping of the mechanical oscillator with decay rate $\gamma$ due to interaction with a mechanical reservoir at temperature $T$. The thermal occupation number of the mechanical bath is denoted by ${\bar n}_{\rm th}=[{\rm exp}(\hbar\omega_m/k_BT)-1]^{-1}$.

\subsection{Master equation in the interaction picture}
To proceed it is convenient to introduce the unitary operator $\hat U_1$ that transforms to a frame rotating at the driving frequency $\omega_L$ for the cavity field
\begin{equation}
\hat U_1 =e^{-i\omega_L\hat a^\dag\hat a t} ,
\end{equation}
and the unitary displacement operator $\hat U_2$ capturing the steady-state mean amplitude of the cavity field resulting from the external pump
\begin{equation}
 \hat U_2 =e^{(\alpha\hat a^\dag -\alpha^* \hat a)},
\end{equation}
with steady-state intracavity amplitude $\alpha$ given by
\begin{equation}
\alpha=\frac{\eta}{-i\Delta_c+\kappa/2} \equiv \sqrt{n_c}.
\end{equation}
Without loss of generality $\alpha$ is here chosen as real by judicious choice of the phase of the pumping rate $\eta$, and $\Delta_c=\omega_L-\omega_c$ is the detuning of the pump laser from the resonance. 
The master equation for the transformed density operator $\bar\rho =  \hat U_2^\dag \hat U_1^\dag \tilde \rho \hat U_1 \hat U_2$ then becomes
\begin{eqnarray}
 \dot{\bar \rho} &=&i\Delta_c[\hat a^\dag\hat a, \bar \rho]-i\omega_m^\prime[\hat b^\dag\hat b, \bar \rho]-ig_0n_c[\hat b^{\dag 2}+\hat b^2, \bar \rho] \nonumber \\
 &&-ig[(\hat a+\hat a^\dag)(\hat b^{\dag}+\hat b)^2, \bar \rho] -ig_0[\hat a^\dag\hat a(\hat b^{\dag}+\hat b)^2, \bar \rho] \nonumber \\
 &&+\frac{\kappa}{2}{\cal D}[\hat a]\bar\rho +\frac{\gamma}{2}{\bar n}_{\rm th}{\cal D}[\hat b^\dag]\bar\rho+\frac{\gamma}{2}({\bar n}_{\rm th}+1){\cal D}[\hat b]\bar\rho, \label{master_temp}
\end{eqnarray}
where $\omega_m^\prime = \omega_m + 2g_0n_c$ is the shifted frequency of the mechanical oscillator, and $g=g_0\sqrt{n_c}$.  This frequency shift proportional to the intracavity photon number comes from the quadratic optomechanical interaction, as opposed to the displacement of the mechanical equilibrium position that rises for the case of linear optomechanical coupling. 

In the regime in which the mean cavity photon number $n_c$ is much larger than the photon fluctuations, the fifth term on the right-hand-side of Eq.~(\ref{master_temp}) may be neglected:  This term is a factor $1/n_c$ smaller than the third term and a factor $1/\sqrt{n_c}$ smaller than the fourth term, these also arising from the quadratic interaction. Following this approximation leads to an optomechanical interaction that is linear in the cavity field operators.

In order to investigate the mechanics in the deep quantum regime, we proceed by assuming that the external pump is red-detuned by twice the effective mechanical frequency, $\Delta_c=-2\omega_m^\prime$.  Then a further simplification follows by invoking the rotating-wave approximation in the interaction picture implemented by the unitary transformation $\hat U_3 = e^{i(\Delta_c\hat a^\dag\hat a-\omega_m^\prime\hat b^\dag\hat b)t}$, and the resulting master equation becomes
\begin{eqnarray}
 \dot{\rho} &=&
-ig[\hat a^\dag\hat b^2+\hat b^{\dag 2}\hat a, \rho] \nonumber \\
 &&+\frac{\kappa}{2}{\cal D}[\hat a]\rho +\frac{\gamma}{2}{\bar n}_{\rm th}{\cal D}[\hat b^\dag]\rho+\frac{\gamma}{2}({\bar n}_{\rm th}+1){\cal D}[\hat b]\rho, \label{master}
\end{eqnarray}
where $\rho = \hat U_3^{\dag}\bar\rho\hat U_3$.  We note that the third term on the right-hand-side of Eq.~(\ref{master_temp}) has been neglected on the basis that it is off-resonant and counter-rotating if $g_0n_c \ll \omega_m^\prime$, and we have checked numerically that this term is indeed negligible in the weak coupling regime. Physically the Hamiltonian representing the Schr{\"o}dinger evolution in Eq.~(\ref{master}) reads
\begin{equation}
 \hat H = \hbar g(\hat a^\dag\hat b^2+\hat b^{\dag 2}\hat a),
\end{equation}
and is identical to the interaction picture Hamiltonian describing a parametric amplifier in quantum optics and is well-known to generate two photons in the subharmonic mode ($\hat b$) destroying a photon in the pump mode ($\hat a$)~\cite{Milburn_book1}. It is thus expected that two phonons of the mechanics can be destroyed by creating a single photon which is eventually leaked out the optical resonator by the cavity field dissipation at rate $\kappa$. 

\subsection{Reduced density operator for the mechanics}
In the regime where cavity dissipation is the dominant source of damping, the state of the cavity field tends to approach to a coherent state in a timescale of $1/\kappa$ and thus the density operator describing the optomechanical system can be approximated as a product state
\begin{equation}
 \rho(t) \approx \rho_o(t) \otimes \rho_m(t),
\end{equation}
where $\hat \rho_o$ is the reduced density operator for the cavity field and $\hat \rho_m$ is the reduced density operator for the mechanics. One should keep in mind that on a timescale slower than $1/\kappa$, the dynamics of the optomechanical system is dependent of that of the mechanical oscillator whereas the dynamics of the cavity field is instantaneously followed by that of the mechanics due to the fast dissipation of the cavity field. Specifically, the reduced density operator for the cavity field describes the vacuum state, $\hat \rho_o= (|0\rangle\langle 0|)_o$ in that we are already in the displaced field picture. 

In order to properly eliminate the reduced density operator for the cavity field and to derive the effective master equation for the mechanical oscillator, we follow the approach used for eliminating the density operator for the pump mode of a parametric amplifier in quantum optics or for the cavity field in cavity QED, see e.g. ~\cite{Carmichael_book}. The dynamics of the reduced density operator for the mechanics is then described by the effective master equation
\begin{eqnarray}
 \frac{d\rho_m}{dt} &=&\frac{\Gamma_{\rm opt}}{2}{\cal D}[\hat b^2]\rho_m \nonumber\\
 &&+\frac{\gamma}{2}{\bar n}_{\rm th}{\cal D}[\hat b^\dag]\rho_m +\frac{\gamma}{2}({\bar n}_{\rm th}+1){\cal D}[\hat b]\rho_m,
\end{eqnarray}
where $\Gamma_{\rm opt}$ is the nonlinear optomechanical damping rate given by 
\begin{equation}
\Gamma_{\rm opt} = \frac{8g^2}{\kappa}.
\end{equation}
Note that this rate is identical to the maximum value of the optomechanical damping rate for $\kappa \ll \omega_m$~\cite{Review1}. The first term on the right-hand-side of the effective master equation accounts for two-phonon damping of the mechanical oscillator and the damping rate is proportional to the cavity photon number, indicating that the mechanical oscillator experiences the optical reservoir at zero temperature through the cavity field. In other words, the intracavity photon number can be used as a control parameter for the nonlinear optomechanical coupling strength of the mechanics to optical reservoir. That is, the dynamics and steady-state properties of mechanical oscillator are affected by two independent heat baths: The optical bath at zero temperature via two phonon processes and mechanical bath at finite temperature via one phonon processes. 

It is convenient to scale time to the inverse of the mechanical decay rate, $\tau=\gamma t$, in terms of which the effective master equation for the mechanics then becomes
\begin{eqnarray}
 \frac{d \rho_m}{d\tau} &=&\frac{C}{2}{\cal D}[\hat b^2]\rho_m \nonumber \\
 &&+\frac{1}{2}{\bar n}_{\rm th}{\cal D}[\hat b^\dag]\rho_m +\frac{1}{2}({\bar n}_{\rm th}+1){\cal D}[\hat b]\rho_m,
\end{eqnarray}
where the multiphoton cooperativity $C$ is given by
\begin{equation}
 C=\frac{\Gamma_{\rm opt}}{\gamma}=\frac{8g^2}{\gamma\kappa} .
\end{equation}
The multiphoton cooperativity is dimensionless and is a measure of the relative coupling strengths of the mechanical oscillator to the cavity-filtered optical bath and mechanical heat bath. Large cooperativity compared to the thermal occupation number $\bar n_{\rm th}$ indicates that mechanical oscillator is more influenced by the optical bath than the mechanical bath and the dynamics of the mechanics is highly nonlinear. 


\section{Results}\label{sec:Results}
We next turn to the analytic solution of the master equation for the mechanics in the high and low temperature regimes.  For this purpose we employ well known phase-space methods that we now discuss briefly as applied to our case.
\subsection{Phase-space methods}
As is well-known, a nonlinear quantum mechanical problem can be mapped into a classical stochastic process by an appropriate phase space representation. We proceed to derive the equation of motion for the mechanical system in the complex $P$ representation. Expanding the density operator for the mechanics as 
\begin{equation}
 \rho_m = \int \frac{|\mu\rangle\langle \nu^*|}{\langle \nu^*|\mu\rangle}P(\mu, \nu) {\rm d}\mu{\rm d}\nu,
\end{equation}
and making use of the quantum correspondence appropriate for the complex $P$ representation~\cite{Generalized_P_represenation}  
\begin{eqnarray}
 \hat b \rho_m &\leftrightarrow& \mu P(\mu, \nu), \\
 \hat b ^\dag  \rho_m &\leftrightarrow& \left(\nu-\frac{\partial}{\partial \mu}\right) P(\mu, \nu), \\
  \rho_m \hat b ^\dag &\leftrightarrow& \nu P(\mu, \nu), \\
  \rho_m \hat b &\leftrightarrow& \left(\mu-\frac{\partial}{\partial \nu}\right) P(\mu, \nu),
\end{eqnarray}
the master equation for the mechanics takes the form of the Fokker-Planck equation
\begin{eqnarray}
 \frac{dP({\bf \chi})}{d\tau}&=& -\sum_i \frac{\partial}{\partial \chi_i}[A({\bf\chi})]_i P({\bf \chi}) \nonumber \\
 &&+\frac{1}{2}\sum_{i,j}\frac{\partial}{\partial \chi_i}\frac{\partial}{\partial \chi_j}[D({\bf\chi})]_{i,j}P({\bf \chi}), \label{FP_equation}
\end{eqnarray}
where ${\bf \chi} =(\mu, \nu)^T$, the drift vector $A({\bf\chi})$ is given by
\begin{equation}
A({\bf\chi}) = 
\begin{pmatrix}
  -\frac{1}{2}\mu -C\nu\mu^2  \\
  -\frac{1}{2}\nu -C\mu\nu^2 
 \end{pmatrix},
\end{equation}
and the diffusion matrix $D({\bf\chi})$ is 
\begin{equation}
D({\bf\chi}) = 
\begin{pmatrix}
  -C\mu^2 & {\bar n}_{\rm th}  \\
  {\bar n}_{\rm th} & -C\nu^2
 \end{pmatrix}.
\end{equation}
We remark that Eq.~(\ref{FP_equation}) is identical to that of the complex $P$ distriubution for single-mode optical field in a cavity that involves cubic-nonlinear dispersive medium~\cite{Drummond_bistability}. We further note that there are two diffusion sources for the complex $P$ distribution function, thermal fluctuations due to mechanical heat bath represented by the off-diagonal elements of the diffusion matrix, and additional quantum fluctuations due to the optomechanical interaction represented by the diagonal elements.   Given the steady-state complex distribution function $P_s$ all normally-ordered steady-state moments can be calculated as 
\begin{equation}
 \langle (\hat b^{\dag})^n(\hat b)^{n'} \rangle_{\rm ss} = \int {\rm d}\mu~(\mu^*)^n(\mu)^{n'}P_s(\mu, \mu^*). \label{moment}
\end{equation}

\subsection{High temperature regime}
In the regime where the thermal fluctuations are the dominant source of diffusion, $\bar{n}_{\rm th}\gg C$, we are able to neglect quantum fluctuations resulting from the optomechanical interaction so that the diffusion matrix can be approximated as  
\begin{equation}
D({\bf\chi}) \approx 
\begin{pmatrix}
  0 & {\bar n}_{\rm th}  \\
  {\bar n}_{\rm th} & 0
 \end{pmatrix}.
\end{equation}
Then setting the left-hand-side of Eq.~(\ref{FP_equation}) to zero for steady-sate, and employing the usual potential condition~\cite{Milburn_book1}, the distribution function is readily found as
\begin{equation}
 P_s(\mu, \nu) = {\cal N} \exp\left({-\frac{1}{ \bar{n}_{\rm th}}\mu\nu}\right) \exp\left({-\frac{C}{\bar{n}_{\rm th}}\mu^2\nu^2}\right), 
\end{equation}
where ${\cal N}$ is a normalization constant. Note that this complex $P$ distribution is bounded and well behaved in the domain in which $\nu=\mu^*$, namely, the Glauber-Sudarshan $P$ representation can be used~\cite{Drummond_para_amp}. The corresponding Glauber-Sudarshan $P$ distribution becomes
\begin{equation}
  P_s(\mu, \mu^*) = {\cal N} \exp\left({-\frac{1}{ \bar{n}_{\rm th}}|\mu|^2}\right) \exp\left({-\frac{C}{\bar{n}_{\rm th}}|\mu|^4}\right), \label{GS_distribution}
\end{equation}
From this result we see that for $C\ll1$ the Glauber-Sudarshan $P$ distribution approaches that for a thermal mixture with occupation number $\bar n_{th}$
\begin{equation}
  P_s(\mu, \mu^*) \approx \frac{1}{{\bar n}_{\rm th}\pi} \exp\left({-\frac{1}{ \bar{n}_{\rm th}}|\mu|^2}\right), 
\end{equation}
as expected in the limit of small multiphoton cooperativity~\cite{Quantum_noise}. On the other hand the Glauber-Sudarshan $P$ distribution can be approximated as
\begin{equation}
   P_s(\mu, \mu^*) \approx \frac{2}{\pi^{3/2}}\sqrt{\frac{C}{\bar{n}_{\rm th}}}\exp\left({-\frac{C}{\bar{n}_{\rm th}}|\mu|^4}\right),
\end{equation} 
in the limit of large multiphoton cooperativity $C\gg1$.

\begin{figure}[]
\includegraphics[width=0.48 \textwidth]{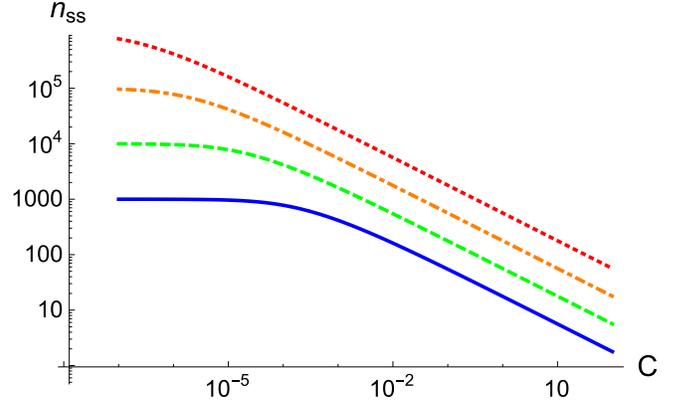}
\caption{\label{fig:cooling}   (Color online) Steady-state mean phonon number of the mechanics as a function of the multiphoton cooperativity $C$ for different thermal occupation numbers $\bar n_{\rm th}$; $\bar n_{\rm th}=10^6$ (red dotted line), $\bar n_{\rm th}=10^5$ (orange dot-dashed line), $\bar n_{\rm th}=10^4$ (green dashed line), $\bar n_{\rm th}=10^3$ (blue solid line).}
\end{figure}

In Fig.~\ref{fig:cooling} we plot the steady-state mean phonon number obtained from Eq.~(\ref{moment})
\begin{equation}
 \langle\hat b^\dag \hat b\rangle_{\rm ss}\equiv n_{\rm ss}= -\frac{1}{2C}+\sqrt{\frac{\bar{n}_{\rm th}}{\pi C}}\frac{\exp\left({-\frac{1}{4C\bar{n}_{\rm th}}}\right)
}{{\rm erfc}\left(\sqrt{\frac{1}{4C\bar{n}_{\rm th}}}\right)},
\end{equation}
versus the multiphoton cooperativity $C$ for different thermal phonon numbers $\bar n_{th}$.  Here ${\rm erfc}(x)=1-{\rm erf}(x)$ is the complementary error function.  The results show that the mechanics, in a thermal state of mean occupation number ${\bar n}_{\rm th}$ at low multiphoton cooperativity, is cooled down as the multiphoton cooperativity $C$ is increased.  Indeed the steady-state mean phonon number approaches 
\begin{equation}
 n_{\rm ss} \approx  \sqrt{\frac{\bar n_{\rm th}}{{\pi C}}}.
\end{equation}
in the limit of large multiphoton cooperativity $C\gg 1$.

\begin{figure}[]
\includegraphics[width=0.48 \textwidth]{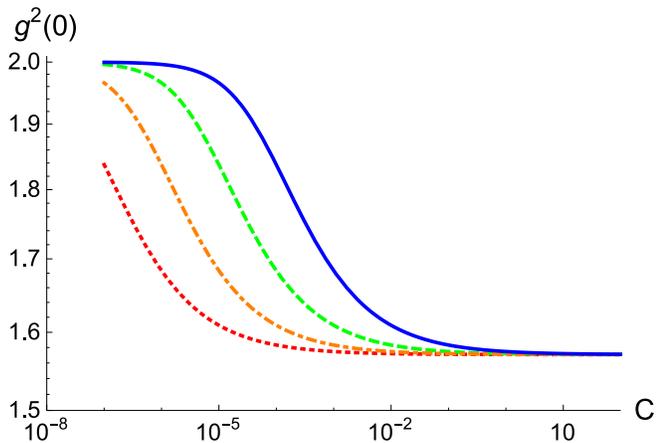}
\caption{\label{fig:correlation}   (Color online) Steady-state second-order correlation function $g^{(2)}(0)$ as a function of the multiphoton cooperativity $C$ for different thermal occupation numbers $\bar n_{\rm th}$; $\bar n_{\rm th}=10^6$ (red dotted line), $\bar n_{\rm th}=10^5$ (orange dot-dashed line), $\bar n_{\rm th}=10^4$ (green dashed line), $\bar n_{\rm th}=10^3$ (blue solid line).}
\end{figure}

\begin{figure}[]
\includegraphics[width=0.48 \textwidth]{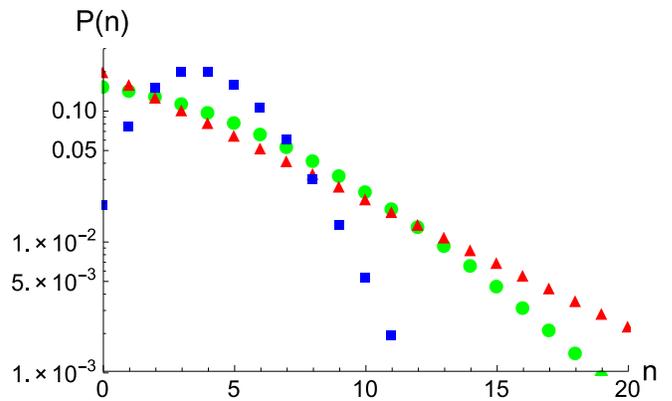}
\caption{\label{fig:number_distribution}   (Color online) Steady-state phonon number distribution $P(n)$ of the mechanical oscillator (green circles) for ${\bar n}_{\rm th}= 10^4,~C= 10^2$. For comparison, the phonon number distributions of the mechanics in a thermal state (red triangles) and a coherent state (blue squares) with the same  the mean phonon number are shown.}
\end{figure}

To probe further we calculate the second-order correlation function defined as
\begin{equation}
 g^{(2)}(0) \equiv \frac{\langle\hat b^{\dag 2}\hat b^2\rangle_{\rm ss}}{\langle\hat b^{\dag}\hat b\rangle_{\rm ss}^2} ,
 \end{equation}
this being plotted in Fig.~\ref{fig:correlation} as a function of the multiphoton cooperativity for different thermal occupation numbers. This figure makes clear that in the regime where $C\ll1$ the second-order correlation function $g^{(2)}(0)$ becomes $2$, a feature of a thermal state.  On the other hand, $g^{(2)}(0)$ approaches $\pi/2$ for large multiphoton cooperativity, indicating that the steady-state of the mechanics is chaotic. This tendency stems from the fact that the linear thermal fluctuations overwhelm the nonlinear two-phonon optomechanical cooling. As a result, the phonon distribution is always bunched in the high temperature regime, and the variance of the phonon number distribution for the mechanical oscillator is in-between those of the mechanics in a thermal equilibrium and a coherent state with the same mean phonon number.  This is illustrated in Fig.~\ref{fig:number_distribution} which shows the steady-state phonon number distribution $P(n)$ of the mechanical oscillator (green circles) for ${\bar n}_{\rm th}= 10^4,~C= 10^2$, along with the cases of a thermal state (red triangles) and a coherent state (blue squares) for comparison. 
\subsection{Low temperature regime}
\begin{figure}[]
\includegraphics[width=0.48 \textwidth]{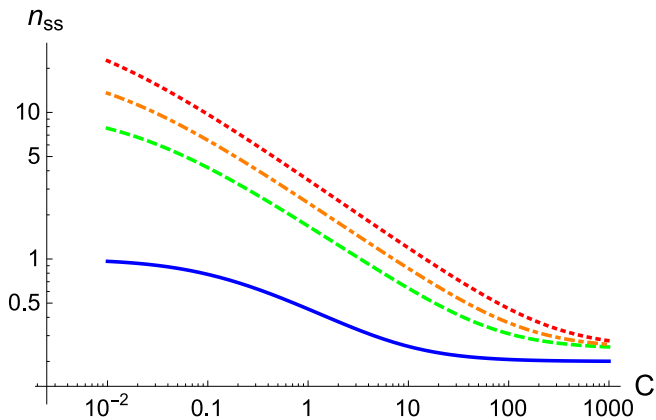}
\caption{\label{fig:phonon_quantum} (Color online) Steady-state mean phonon number $n_{\rm ss}$ of the mechanical oscillator as a function of multiphoton cooperativitiy $C$ in the low temperature regime: ${\bar n_{\rm th}}=1$ (blue solid line), ${\bar n_{\rm th}}=10$ (green dashed line), ${\bar n_{\rm th}}=20$ (orange dot-dashed line), and ${\bar n_{\rm th}}=40$ (red dotted line).}
\end{figure}

\begin{figure}[]
\includegraphics[width=0.48 \textwidth]{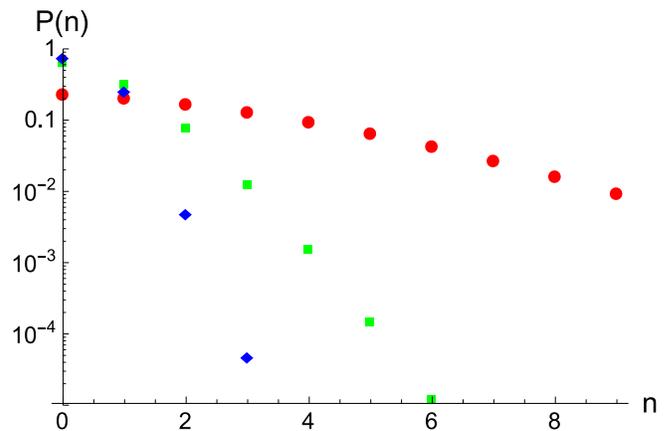}
\caption{\label{fig:population_quantum} (Color online) Steady-state phonon  distributions $P(n)$ of the mechanical oscillator with different multiphoton cooperativity at the same temperature, ${\bar n_{\rm th}}=20$: $C=1$ (red circles), $C=41$ (green squares), and $C=1000$ (blue rombi).}
\end{figure}

In order to explore the possibility of an antibunched phonon field, a key signature that the mechanical system is in a truly quantum state, we proceed to examine the low temperature regime. We have obtained the steady-state complex $P$ distribution following the procedures outlined in Ref.~\cite{Nonlinear_damping}, but for the sake of clarity in presentation we relegate the details to the Appendix and concentrate on the results here.  Specifically, we find that the complex $P$ distribution is given by
\begin{eqnarray}
 P_s(\mu, \nu) &=& \frac{2Ae^{2\mu\nu}}{(1+2\bar{n}_{\rm th}-C)\mu\nu}
 {}_2F_1\left(1,1;\tfrac{1+2\bar{n}_{\rm th}}{C};\tfrac{\bar{n}_{\rm th}}{C\mu\nu}\right) \nonumber \\
  &&+\frac{2Ae^{2\mu\nu}}{\bar{n}_{\rm th}}\sum_{r=1}^{\infty}\frac{(-2\mu\nu)^r}{rr!}\times \nonumber \\
 &&{}_2F_1\left(1,2+r-\tfrac{1+2\bar{n}_{\rm th}}{C};1+r;\tfrac{C\mu\nu}{\bar{n}_{\rm th}}\right), \label{distribution_quantum}
\end{eqnarray} 
where ${}_2F_1(a,b;c;z)$ is the hypergeometric function.  The corresponding expression for the steady-state mean phonon number of the mechanics is given by Eq.~(\ref{mean_phonon_quantum}), and is plotted in Fig.~\ref{fig:phonon_quantum} as a function of the multiphoton cooperativity $C$, and for a variety of thermal occupation numbers.  These results show that the mechanics is cooled down near the motional ground state in the regime where $C\gg1$. In this regime the optomechanical two-phonon damping is dominant so that only the ground and first-excited states are significantly populated (see steady-state phonon distribution indicated by blue rhombi in Fig.~\ref{fig:population_quantum}). Furthermore, the population of the mechanics in the first-excited state tends to increase with increasing temperature. These results are in accordance with the numerical calculations based on the Fock-state representation~\cite{Quadratic1}.  

\begin{figure}[]
\includegraphics[width=0.48 \textwidth]{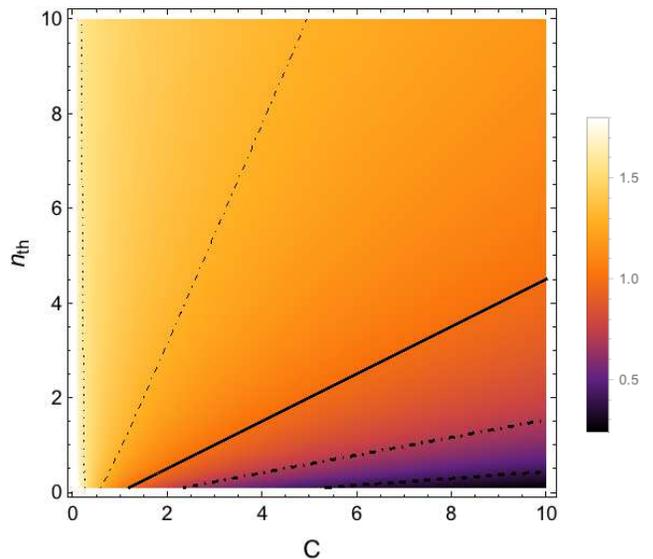}
\caption{\label{fig:g2_quantum} (Color online) Steady-state second-order correlation function $g^{(2)}(0)$ of the mechanical oscillator as a function of both the multiphoton cooperativitiy and thermal occupation number.  The various lines show contours of constant $g^{(2)}(0)$: $g^{(2)}(0)=1.6$ (dotted line), $g^{(2)}(0)=1.3$ (dot-dashed line), $g^{(2)}(0)=1.0$ (thick solid line), $g^{(2)}(0)=0.7$ (thick dot-dashed line), and $g^{(2)}(0)=0.4$ (thick dashed line).}
\end{figure}

The expression for the second-order correlation function $g^{(2)}(0)$ of the mechanics is given by Eq.~(\ref{g2_quantum}). Fig.~\ref{fig:g2_quantum} shows a color coded plot of the second-order correlation function of the mechanical oscillator as a function of both the mutiphoton cooperativity $(C)$ and the thermal occupation number $(\bar n_{th})$.  The plot reveals that the phonon distribution of the mechanics is antibunched $(g^{(2)}(0)<1)$ when  $C> 2{\bar n}_{\rm th}+1$, whereas it is bunched $(g^{(2)}(0)>1)$ when $C<2{\bar n}_{\rm th}+1$.  Physically, the mechanics tends to experience one phonon absorption and emission processes, and its phonon distribution is superpoissonian, if the mechanical thermal and quantum noise sources are dominant, $C<2{\bar n}_{\rm th}+1$. However, in the regime where the optomechanical coupling is stronger than thermal decoherence, $C> 2{\bar n}_{\rm th}+1$, the mechanics has a tendency to experience two-phonon absorption and emission processes and its phonon distribution becomes antibunched. As expected, when $C= 2{\bar n}_{\rm th}+1$ the steady-state of the mechanical oscillator becomes a coherent state with a mean phonon number 
\begin{equation}
 n_{\rm ss}= \frac{{\bar n}_{\rm th}}{2{\bar n}_{\rm th}+1},
\end{equation}
and the second-correlation function becomes unity.  This situation is indicated by the thick solid line in Fig.~\ref{fig:g2_quantum}: Regions of parameter space above this line yield steady-state bunching whereas below this line antibunching is realized.  Fig.~\ref{fig:population_quantum} shows three representative plots of the phonon number distributions of the mechanics indicating that only the ground and first-excited states are significantly populated if $C\gg{\bar n}_{\rm th}+1$ (blue rhombi), the distribution being Poissonian if $C=2{\bar n}_{\rm th}+1$ (green circles), and the distribution becoming nearly exponential if $C\ll2{\bar n}_{\rm th}+1$ (red circles).

We finish by noting that in the regime where the mechanical heat bath is at zero temperature
thermal effects are completely negligible compared to the quantum fluctuations, $\bar n_{\rm th}=0$, and the diffusion matrix $D({\bf\chi})$ reads
\begin{equation}
D({\bf\chi})= 
\begin{pmatrix}
  -C\mu^2 & 0  \\
  0 & -C\nu^2
 \end{pmatrix}.
\end{equation}
This situation was previously studied extensively in the context of quantum optics~\cite{Nonlinear_damping} and the steady-state complex $P$ distribution is given by
\begin{eqnarray}
 P_s(\mu, \nu) =\frac{1-\frac{1}{C}}{2\pi^2}e^{2\mu\nu}\sum_{r=0}^{\infty} \frac{(-2\mu\nu)^{r-1}}{(r+1-\frac{1}{C})r!}.
\end{eqnarray} 
In this case the mechanical oscillator is coupled to an optical reservoir at zero temperature by the nonlinear optomechanical coupling, and is also coupled to the mechanical heat bath at zero temperature by the intrinsic linear interaction.  Then the steady-state of the mechanical oscillator is the motional ground state, as expected, and thus the mean phonon number $n_{\rm ss}=0$ and the second-order correlation function $g^{(2)}(0)=0$~\cite{Nonlinear_damping}. 

\section{Summary and conclusions} \label{Conclusions}
We have analytically investigated the steady-state of a vibrating membrane coupled to a single-mode optical field via a quadratic optomechanical interaction, and in the weak coupling limit.  The mechanics was shown to experience an effective cubic nonlinearity in the limit that the cavity dissipation rate is much larger than both the optomechanical coupling and mechanical damping rates, allowing for adiabatic elimination of the cavity field.  Our key result is that the steady-state phonon field is chaotic if the multiphoton cooperativity obeys $C<2{\bar n_{\rm th}}+1$ whereas it antibunched if $C>2{\bar n_{\rm th}}+1$. 

There are of course barriers to realizing antibunching of a phonon field, but recent developments make this more feasible.  The requirement of large optomechanical cooperativity has been realized in high-frequency optomechanical oscillators~\cite{Cooperativity}, with a quoted maximum value of $146,000$.  In addition, the demonstration of a Hanbury-Brown-Twiss type experiment ~\cite{Phonon_counting} for a phonon field in a nanomechanical resonator paves the way to measuring the second-order correlation.  Thus our calculation opens the door to control of the second-order correlation of the mechanical oscillator in the weak coupling regime, and the observation of phonon antibunching.

\acknowledgments
This work is supported by the Korea National Reserach Foundation (NRF) NRF-2015R1C1A1A01052349. 

\appendix

\section{Steady-state complex $P$ distribution in the low temperature regime}\label{appendix A}

In order to find the steady-state complex $P$ distribution of the mechanics, we follow the procedures outlined in Ref.~\cite{Nonlinear_damping}.   Equation~(\ref{FP_equation}) can be written as
\begin{eqnarray}
  \frac{dP}{d\tau}&=& \frac{\partial}{\partial\mu}\left[\frac{\mu}{2}+C\mu^2\nu-\frac{C}{2}\frac{\partial}{\partial\mu}\mu^2+\frac{\bar{n}_{\rm th}}{2}\frac{\partial}{\partial\nu}\right]P\nonumber \\
  &&+ \frac{\partial}{\partial\nu}\left[\frac{\nu}{2}+C\nu^2\mu-\frac{C}{2}\frac{\partial}{\partial\nu}\nu^2+\frac{\bar{n}_{\rm th}}{2}\frac{\partial}{\partial\mu}\right]P.
\end{eqnarray}
The steady-state complex $P$ distribution can in general be obtained from 
\begin{eqnarray}
\left[\frac{\mu}{2}+C\mu^2\nu-\frac{C}{2}\frac{\partial}{\partial\mu}\mu^2+\frac{\bar{n}_{\rm th}}{2}\frac{\partial}{\partial\nu}\right]P_s&=&f(\nu), \label{b1}\\
\left[\frac{\nu}{2}+C\nu^2\mu-\frac{C}{2}\frac{\partial}{\partial\nu}\nu^2+\frac{\bar{n}_{\rm th}}{2}\frac{\partial}{\partial\mu}\right]P_s&=&g(\mu), \label{b2} 
\end{eqnarray}
where $f(\nu)$ and $g(\mu)$ must satisfy generalized potential conditions~\cite{Nonlinear_damping}. To find the form of these functions we write $P_s(\mu, \nu)$ as
\begin{equation}
 P_s(\mu, \nu) = \frac{Q(\mu, \nu)}{(C\mu\nu-\bar{n}_{\rm th})^2},
\end{equation}
then Eqs.~(\ref{b1}) and (\ref{b2}) can be written as
\begin{eqnarray}
 \frac{\partial R(\mu, \nu)}{\partial \mu} &=& e^{-2\mu\nu}(C\mu\nu-\bar{n}_{\rm th})^{1-\frac{1+2\bar{n}_{\rm th}}{C}}F(\mu, \nu), \\
  \frac{\partial R(\mu, \nu)}{\partial \nu} &=& e^{-2\mu\nu}(C\mu\nu-\bar{n}_{\rm th})^{1-\frac{1+2\bar{n}_{\rm th}}{C}}G(\mu, \nu),
\end{eqnarray}
where we define for typographical simplicity,
\begin{eqnarray}
 R(\mu, \nu) &=& e^{-2\mu\nu}(C\mu\nu-\bar{n}_{\rm th})^{-\frac{1+2\bar{n}_{\rm th}}{C}}Q(\mu, \nu), \\
 F(\mu, \nu) &=& -2\frac{C\nu^2 f(\nu)+\bar{n}_{\rm th} g(\mu)}{C\mu\nu+\bar{n}_{\rm th}}, \\
 G(\mu, \nu) &=& -2\frac{C\mu^2 g(\mu)+\bar{n}_{\rm th} f(\nu)}{C\mu\nu+\bar{n}_{\rm th}}.
\end{eqnarray}
The generalized potential condition
\begin{equation}
  \frac{\partial^2 R(\mu, \nu)}{\partial \nu\partial \mu}=  \frac{\partial^2 R(\mu, \nu)}{\partial \mu\partial \nu}
\end{equation}
can be written as
\begin{eqnarray}
\frac{\partial}{\partial \nu} \left[e^{-2\mu\nu}(C\mu\nu-\bar{n}_{\rm th})^{1-\frac{1+2\bar{n}_{\rm th}}{C}}F(\mu, \nu)\right]  \nonumber \\
= \frac{\partial}{\partial \mu}\left[e^{-2\mu\nu}(C\mu\nu-\bar{n}_{\rm th})^{1-\frac{1+2\bar{n}_{\rm th}}{C}}G(\mu, \nu)\right] \label{potential_condition}.
\end{eqnarray}
Equation~(\ref{potential_condition}) is satisfied for 
\begin{eqnarray}
 f(\nu) &=& \frac{A}{\nu}, \\
 g(\mu) &=& \frac{A}{\mu},
\end{eqnarray}
where $A$ is a constant.
Thus, the steady-state complex $P$ distribution is given by after some algebra
\begin{equation}
 P_s(\mu, \nu) = e^{2\mu\nu}(C\mu\nu-\bar{n}_{\rm th})^{\frac{1+2\bar{n}_{\rm th}}{C}-2}[B+AI(\mu, \nu)],
\end{equation}
where $B$ is a constant of integration and $I(\mu, \nu)$ is the indefinite integral
\begin{equation}
 I(\mu, \nu) = -2\int {\rm d}\mu~\frac{e^{-2\mu\nu}}{\mu}(C\mu\nu-\bar{n}_{\rm th})^{1-\frac{1+2\bar{n}_{\rm th}}{C}}. \label{I}
\end{equation}
This integral may be calculated using a power-series expansion of the exponential function and the resulting steady-state complex $P$ distribution reads
 \begin{eqnarray}
 P_s(\mu, \nu) &=& Be^{2\mu\nu}(C\mu\nu-\bar{n}_{\rm th})^{\frac{1+2\bar{n}_{\rm th}}{C}-2} \nonumber\\
 &&+\frac{2Ae^{2\mu\nu}}{(1+2\bar{n}_{\rm th}-C)\mu\nu}
 {}_2F_1\left(1,1;\tfrac{1+2\bar{n}_{\rm th}}{C};\tfrac{\bar{n}_{\rm th}}{C\mu\nu}\right) \nonumber \\
  &&+\frac{2Ae^{2\mu\nu}}{\bar{n}_{\rm th}}\sum_{r=1}^{\infty}\frac{(-2\mu\nu)^r}{rr!}\times \nonumber \\
 &&{}_2F_1\left(1,2+r-\tfrac{1+2\bar{n}_{\rm th}}{C};1+r;\tfrac{C\mu\nu}{\bar{n}_{\rm th}}\right).
\end{eqnarray} 
It should be noted that the two constants $A$ and $B$ are chosen from the normalization condition and the requirement that the phonon number distribution be nonnegative. Using the complex $P$ distribution function, all normal-ordered moments in the steady state can be obtained from
\begin{equation}
 \langle (\hat b^{\dag})^n(\hat b)^{n'} \rangle_{\rm ss} = \int {\rm d}\mu{\rm d}\nu~(\nu)^n(\mu)^{n'}P_s(\mu, \nu). \label{moment_complex_P}
\end{equation}
Making the change of variables
\begin{eqnarray}
 N&=& \mu\nu, \\
 z&=& \mu,
\end{eqnarray}
and choosing a circular contour around the origin for the $z$ line integral, and a Hankel contour for the $N$ line integral~\cite{Math_book2}, one can find the normalization condition, the mean phonon number, the second-order correlation, and so on from Eq.~(\ref{moment_complex_P}).
The normalization condition reads
\begin{eqnarray}
\frac{1}{4\pi^2}&=& -B\frac{e^{\frac{2{\bar n}_{\rm th}}{C}}}{2\Gamma\left(2-\frac{1+2{\bar n}_{\rm th}}{C}\right)} \left(\frac{C}{2}\right)^{\frac{1+2{\bar n}_{\rm th}}{C}-2} \nonumber \\
 &&-A\frac{2\Gamma\left(\frac{1+2{\bar n}_{\rm th}}{C}\right)}{1+2{\bar n}_{\rm th}-C}\sum_{k=0}^{\infty}\frac{(2{\bar n}_{\rm th}/C)^k}{\Gamma\left(\frac{1+2{\bar n}_{\rm th}}{C}+k\right)}.
\end{eqnarray}
The populations of the $m$-th number state  are given by
\begin{eqnarray}
 P_{m} &=& -B\sum_{k=0}^{m}\frac{4\pi^2e^{\frac{{\bar n}_{\rm th}}{C}}C^{\frac{1+2{\bar n}_{\rm th}}{C}-2} \left(\frac{{\bar n}_{\rm th}}{C}\right)^{m-k}}{\Gamma(m-k+1)\Gamma(k+1)\Gamma\left(2-k-\frac{1+2{\bar n}_{\rm th}}{C}\right)}  \nonumber \\
 &&-A\frac{8\pi^2\Gamma\left(\frac{1+2{\bar n}_{\rm th}}{C}\right)}{(1+2{\bar n}_{\rm th}-C)m!}\times \nonumber \\
 &&\sum_{k=m}^{\infty}\frac{\Gamma(k+1)({\bar n}_{\rm th}/C)^k}{\Gamma\left(\frac{1+2{\bar n}_{\rm th}}{C}+k\right)\Gamma(k+1-m)}.
\end{eqnarray}
In order for the phonon number distribution to be nonnegative, $B=0$ for $C\neq1+2\bar{n}_{\rm th}$ and $A=0$ for $C=1+2\bar{n}_{\rm th}$ due to the oscillatory behavior of the $\Gamma$ function. 

If $C\neq1+2\bar{n}_{\rm th}$, normalization constant $A$ is given by
\begin{equation}
 A = -\frac{1+2\bar{n}_{\rm th}-C}{8\pi^2\Gamma\left(\frac{1+2{\bar n}_{\rm th}}{C}\right) \displaystyle\sum_{k=0}^{\infty}\frac{(2{\bar n}_{\rm th}/C)^k}{\Gamma\left(\frac{1+2{\bar n}_{\rm th}}{C}+k\right)}}.
\end{equation}
The mean phonon number is given by
\begin{equation}
n_{\rm ss}= \frac{1}{2}\frac{\displaystyle\sum_{k=0}^\infty \frac{k}{\Gamma\left(\frac{1+2{\bar n_{\rm th}}}{C}+k\right)}\left(\frac{2{\bar n_{\rm th}}}{C}\right)^k}{\displaystyle\sum_{k=0}^\infty \frac{1}{\Gamma\left(\frac{1+2{\bar n_{\rm th}}}{C}+k\right)}\left(\frac{2{\bar n_{\rm th}}}{C}\right)^k}, \label{mean_phonon_quantum}
\end{equation}
and the second-order correlation function $g^{(2)}(0)$ is
\begin{equation}
 g^{(2)}(0)= \frac{\displaystyle\sum_{k, k\prime}^\infty \frac{k(k-1)}{\Gamma\left(\frac{1+2{\bar n_{\rm th}}}{C}+k\right)\Gamma\left(\frac{1+2{\bar n_{\rm th}}}{C}+k^\prime\right)}\left(\frac{2{\bar n_{\rm th}}}{C}\right)^{k+k^\prime}}{\displaystyle\sum_{k, k\prime}^\infty \frac{kk^\prime}{\Gamma\left(\frac{1+2{\bar n_{\rm th}}}{C}+k\right)\Gamma\left(\frac{1+2{\bar n_{\rm th}}}{C}+k^\prime\right)}\left(\frac{2{\bar n_{\rm th}}}{C}\right)^{k+k^\prime}}. \label{g2_quantum}
\end{equation}

\end{document}